\documentclass[twocolumn,aps,prl,reprint,groupedaddress]{revtex4}
\usepackage{graphicx}
\usepackage{color}

\newcommand{\be}{\begin{equation}}
\newcommand{\ee}{\end{equation}}

\newcommand{\bea}{\begin{eqnarray}}
\newcommand{\eea}{\end{eqnarray}}

\newcommand{\la}{\langle}
\newcommand{\ra}{\rangle}
\newcommand{\lb}{\left[}
\newcommand{\rb}{\right]}
\newcommand{\lp}{\left(}
\newcommand{\rp}{\right)}

\renewcommand{\vec}[1]{{\bf #1}}

\newcommand{\addJS}[1]{\textcolor{red}{#1}}

\begin{document}
\title{%Disorder-Assisted Electron-Lattice Cooling 
 Disorder-Assisted Electron-Phonon Scattering and Cooling Pathways in Graphene}

\author{Justin C. W. Song$^{1,2}$}
%\email{jcwsong@mit.edu}
\author{Michael Y. Reizer$^3$}
\author{Leonid S. Levitov$^1$}
%\email{levitov@mit.edu}
\affiliation{$^1$ Department of Physics, Massachusetts Institute of Technology, Cambridge, Massachusetts 02139, USA}
\affiliation{$^2$ School of Engineering and Applied Sciences, Harvard University, Cambridge, Massachusetts 02138, USA}
\affiliation{$^3$ 5614 Naiche Road, Columbus, Ohio 43213, USA}

%\date{\today}

\begin{abstract}
We predict that graphene is a unique system where disorder-assisted scattering (supercollisions) dominates electron-lattice cooling over a wide range of temperatures, up to room temperature. This is so because for momentum-conserving electron-phonon scattering  the energy transfer per collision is severely constrained due to a small Fermi surface size.
%The rates for electron-lattice cooling in graphene can be slow due to a small Fermi surface size and the momentum-conserving character of electron-phonon scattering, which severely constrain energy transfer per collision.
% We show that alternative cooling pathways mediated by disorder dominate over momentum-conserving channels over a wide range of temperatures under realistic conditions. The disorder-assisted cooling features sign reversal of temperature dependence 
The characteristic $T^3$ temperature dependence
and power-law cooling dynamics provide clear experimental signatures of this new cooling mechanism. The cooling rate can be changed by orders of magnitude by varying the amount of disorder 
% opening the way for using graphene as a platform in a variety of applications
% providing 
which offers means for a variety of new applications that rely on hot-carrier transport.
\end{abstract} 


\maketitle
A number of interesting and practically useful phenomena arise when slow cooling between electronic and lattice systems results in long-lived hot carriers proliferating over large spatial scales\cite{ziman}. Energy transport and energy harvesting mediated by hot carriers is utilized in a variety of applications (calorimetry, bolometry, infrared and THz detectors, etc.). Thermal decoupling of electrons from the crystal lattice in most materials takes place at temperatures of order a few kelvin\cite{giazotto}. 
In contrast, the rates for electron-lattice cooling in graphene 
are predicted to be very slow in a much wider temperature range\cite{macdonald,wong}.
resulting in new optoelectronic and thermoelectric phenomena as well as other hot-carrier effects\cite{rana,winnerl,song,gabor}.

\begin{figure}
\includegraphics[scale=0.29]{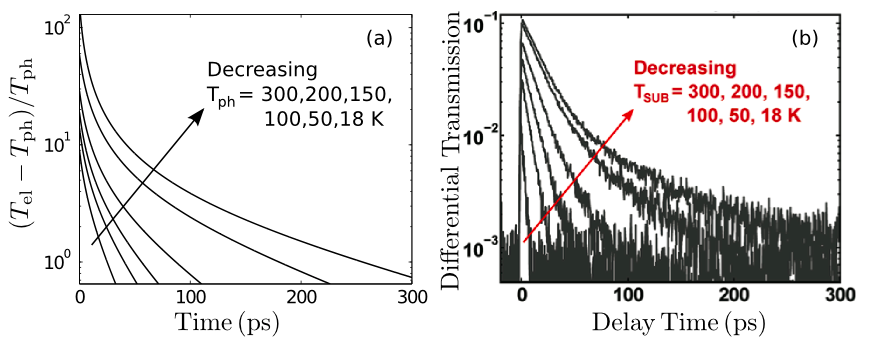} %{bottleneck_fig1_lowres.eps} 
% {bottleneck7.eps} 
\caption{(a) Temperature dynamics 
obtained from Eq.(\ref{eq:Jsuper}) for the lattice temperature values matching those in panel (b). Parameter values used:  doping $\mu = 50 \, {\rm meV}$ and disorder mean free path $k_F\ell = 20$.
% The time scale was given by  $\tau = \alpha/(3AT)$ and $k_F\ell = 20$.  
(b) Carrier dynamics measured using the pump-probe technique for varying substrate temperatures (reproduced from Fig.2(b) of Ref. \cite{rana}).}
\label{compare}
\vspace{-6mm}
\end{figure}

The inefficiency of the standard cooling pathways mediated by optical and acoustic phonons\cite{macdonald,wong} stems from the material properties of graphene. The large value of the optical phonon energy, $\omega_0 = 200\, {\rm meV}$, quenches the optical phonon scattering channel below a few hundred kelvin\cite{footnote:othersys}; a small Fermi surface 
and momentum conservation severely constrain the phase space for acoustic phonon scattering \cite{macdonald,wong}. Given the momentum conserving character of these processes, which renders them inefficient, cooling in graphene can be particularly sensitive to the effects of disorder. In this Letter, we argue that an unconventional, disorder-assisted pathway dominates cooling in a wide range of temperatures, explaining key features of cooling dynamics observed in recent pump-probe measurements \cite{rana} (see Fig. \ref{compare}).

The cooling dynamics reported in Ref.\cite{rana}  features fairly long timescales. The cooling times grow with decreasing  temperature, from $\sim10\,{\rm ps}$ at $300\,{\rm K}$ to $\sim200\,{\rm ps}$ below $50\,{\rm K}$. This is very different from the dependence expected for momentum-conserving scattering by acoustic phonons, where the cooling times are predicted to increase with temperature, reaching a nanosecond scale at room temperature for comparable densities\cite{macdonald,wong}. The observed temperature dependence is also clearly distinct from the  very steep dependence expected for optical phonons, $\tau\sim e^{\hbar\omega_0/k_{\rm B}T}$.
As we show below, the  disorder-assisted cooling mechanism yields slow time scales and a temperature dependence that closely match the observations. In addition, as illustrated in Fig.\ref{compare}, this mechanism explains subtle features such as the prolonged non-exponential regime of cooling dynamics and the saturation  of cooling times at low $T$ manifest in the similarity between the $50\,{\rm K}$ and $18\,{\rm K}$ curves [see Eqs.(\ref{eq:nonmono}),(\ref{eq:1/t})].

The high impact of disorder on cooling can be understood by noting that the momentum-conserving acoustic phonon processes can only dissipate energy in bits much smaller than  $k_BT$. Indeed, since for such processes the phonon momenta are limited by $2k_F$, the maximal energy transfer cannot exceed $2k_BT_{\rm BG} = 2\hbar sk_F$ per scattering event (here $s$ and $k_F$ are the sound velocity and Fermi momentum). The $T_{\rm BG}$ values are {\it a few kelvin} for typical carrier densities, i.e. a small fraction of $k_BT$. In contrast, disorder-assisted scattering allows for arbitrarily large  phonon recoil momentum values. In this case, the entire thermal distribution of phonons can contribute to scattering, resulting in the energy dissipated per scattering of order  $k_BT$ (supercollisions). This provides a dramatic boost to the cooling power.

For this cooling mechanism, 
modeling disorder by short-range scatterers,  we obtain the energy-loss power
% using the short-range scatterer model for disorder, we obtain the energy-loss power
%
\be\label{eq:Jsuper}
\mathcal{J}= A\lp T_{\rm el}^3-T_{\rm ph}^3\rp, \qquad 
%A = \frac{A_0}{k_F\ell}
A=9.62 \frac{g^2\nu^2(\mu) k_{\rm B}^3 }{\hbar k_F\ell}
%, \qquad \lambda = g^2 \nu(\mu)
,
\ee
where $T_{\rm el}$ ($T_{\rm ph}$) the electron (lattice) temperature, $\nu(\mu) $ is the density of states at the Fermi level per one spin/valley flavor, $g$ is the electron-phonon coupling, and $k_{\rm B} T_{\rm el(ph)}\ll\mu$. The enhancement factor for the disorder-assisted cooling over the momentum conserving pathways depends on both disorder and temperature:
\be
\frac{\mathcal{J}}{\mathcal{J}_0}=\frac{0.77}{k_F\ell}\frac{T_{\rm el}^2+T_{\rm el}T_{\rm ph}+T_{\rm ph}^2}{ T_{\rm BG}^2}
,
\label{eq:enhancement}
\ee
[see also Eqs.(\ref{eq:dQ/dt}),(\ref{eq:nonmono})]. At room temperature, $T_{\rm el(ph)}\sim 300\,{\rm K}$, and taking $\mu =100\, {\rm meV}$ ($n \sim10^{12}\,{\rm cm^{-2}}$) we find $T_{\rm el(ph)}/T_{\rm BG}\approx 50$. 
For $k_F\ell =20$,
the enhancement factor ${\mathcal{J}}/{\mathcal{J}_0}$ can be as large as 100 times.

Given the  dominance of the disorder-assisted processes, we predict that cooling in graphene is uniquely sensitive to disorder.$k_{\rm B} T_{\rm el(ph)}\ll\mu$
This sensitivity can account for the wide spread of experimentally measured cooling times\cite{winnerl, rana, dawlaty, plochanka, george}. 
Slow cooling times arise because  $\mathcal{J}$ scales linearly with the disorder concentration, via $1/k_F\ell $, and with carrier density,  $n\sim \nu^2(\mu)$. 
The inverse scaling with $k_F \ell$
is consistent with the trend of cooling becoming faster at higher levels of disorder noted in Ref.\cite{dawlaty}. The sensitivity to disorder can be used as a knob to engineer cooling rates desirable for specific applications.

The enhancement  to phase space may also arise due to  processes of other types\cite{supercollisions}. Recently, Castro et. al. \cite{castro,ochoa} predicted that scattering by  flexural phonons can dominate the resistivity (momentum relaxation) in free standing graphene. 
In contrast, here we are concerned with {\it cooling} 
which is sensitive to both the scattering rate as well as the exchange in energy. For flexural phonons, we find an energy-loss power $T$ dependence resembling that in Eq.(\ref{eq:Jsuper}) but with a greatly diminished prefactor.

In our discussion of cooling we shall implicitly assume that an effective electronic temperature is established quickly via fast carrier-carrier scattering. This is well justified as carrier-carrier scattering occurs on timescales of tens of femtoseconds \cite{rana, george}, far shorter than the timescales $\gtrsim 1\,{\rm ps}$ we are concerned with. Below, we analyze cooling from phonons in the graphene lattice {\it only}. We note that other phonons (particularly, substrate surface phonons\cite{fratini}) may also contribute to cooling\cite{shytov}.

We proceed to analyze the disorder-assisted cooling regime, wherein impurity scattering mediates the exchange of momentum and energy between electron and phonon systems.
The effect of disorder on electron-phonon scattering can be described by the Hamiltonian
\be
\mathcal{H} = \sum_{\vec{k},i}  \psi^\dag_{\vec k,i} H_0(\vec k)\psi_{\vec k,i} + \sum_\vec{q} \omega_{\vec q} b^\dag_\vec{q} b_\vec{q} + \mathcal{H}_{\rm el-ph} + \mathcal{H}_{\rm dis},
\ee
where $H_0= v_F\sigma\cdot\vec k$ is the massless Dirac Hamiltonian, identical for $i=1...N$ spin/valley flavors. The electron-phonon interaction arises from the deformation potential,
% coupling, 
%
\be
\label{eq:H_elph}
\mathcal{H}_{\rm el-ph} = \sum_{\vec q} g \sqrt{\omega_{\vec q}}\big( b_\vec{q} + b^\dag_{-\vec q}\big) n_{\vec q},\quad
g = D/\sqrt{2\rho s^2}
,
\ee
where $D$ is the deformation potential constant and $\rho$ is the mass density of the graphene sheet. 

\begin{figure}
\includegraphics[scale=0.38]{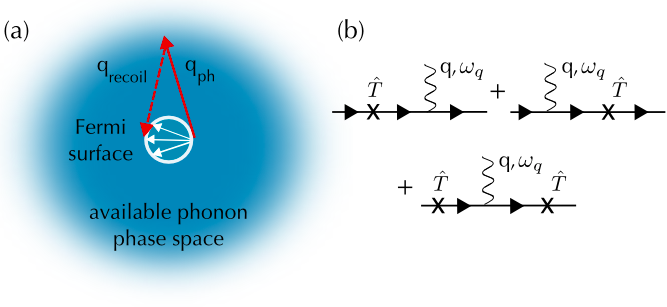} %{bottleneck_fig1_lowres.eps} 
% {bottleneck7.eps} 
\caption{
(a) Kinematics of supercollisions and normal collisions at $T>T_{\rm BG}$. Phonon momenta are constrained by the Fermi surface for normal collisions (white arrows), and totally unconstrained for supercollisions ($q_{\rm ph}$), with the recoil momentum ($q_{\rm recoil}$) transferred to the lattice via disorder scattering or carried away by second phonon. The energy dissipated in supercollisions is much greater than that dissipated in normal collisions.
(b) Feynman diagrams for disorder assisted electron-phonon scattering processes, corresponding to the three terms in Eq.(\ref{matrixelement_M}). 
}
\label{fig2}
\vspace{-6mm}
\end{figure}

The transition probability for the emission and absorption of phonons can be described by Fermi's Golden Rule, 
\be
\label{transition}
W_{\vec{k}',\vec{k}} \!\!=\!\! \frac{2\pi}{\hbar}\sum_{\vec q}\lb |M_+|^2 N_{\omega_{\vec q}}\delta_+ +|M_-|^2 (N_{\omega_{\vec q}}\!\!+1)\delta_-\rb
,
\ee
where $\delta_\pm=\delta( \epsilon_{\vec{k}'}-\epsilon_{\vec{k}}\mp \omega_\vec{q})$, $\vec{q}$ is phonon momentum and $N_{\omega_{\vec q}} = 1/(e^{\beta \omega_\vec{q}}-1)$ is the Bose distribution.
In the absence of disorder, Eq.(\ref{eq:H_elph}) yields the matrix elements 
%are derived from the deformation potential, 
$M_{\pm}^{(0)}=g\sqrt{\omega_{\vec q}}\delta_{\vec k'-\vec k\mp\vec q}$, where the delta function enforces momentum conservation. In the presence of disorder, possible phonon momenta are unconstrained, taking on any value $|\vec q|\lesssim q_T=k_BT/s$ [see Fig.\ref{fig2} (a)]. 

We model the disorder potential as a sum of randomly positioned impurity potentials,
\be\label{eq:H_impurity}
\mathcal{H}_{\rm dis}=  \sum_{ \vec r, i} 
\psi_{i}^\dag(\vec r) U(\vec r) \psi_{i}(\vec r)
,\quad U(\vec r)=\sum_j V(\vec r-\vec r_j)
.
\ee
At low disorder concentration, we can describe disorder-assisted phonon scattering by dressing the electron-phonon vertex with multiple scattering on a single impurity. This gives an expression for the transition matrix elements $M_\pm$ which is exact in the impurity potential:
\be
M_\pm = \la \vec k'|M_{\pm}^{(0)}G\hat T +\hat TGM_{\pm}^{(0)} +  \hat TGM_{\pm}^{(0)}G\hat T |\vec k \ra
% M_{\rm super} = \la \vec k'|M_{\rm ph}G\hat T +\hat TGM_{\rm ph} +  \hat TGM_{\rm ph}G\hat T |\vec k \ra
,
\label{matrixelement_M}
\ee
where $G(\vec p)=\frac1{\epsilon-H_0(\vec p)}$ is the electron Green's function, $\hat T$ is the T-matrix (scattering operator) for a single impurity.
The three terms in Eq.(\ref{matrixelement_M}) account for the cases when impurity scattering occurs before or/and after phonon emission (see Fig. 1 (b)).

As we shall see, the main contribution to cooling will arise from phonons with momenta of order $q_T$. Thus we anticipate that the virtual electron states, described by the Green's functions $G(\vec p)$, are characterized by large momenta $|\vec p|\sim q_T$ which are much greater than $\vec k$, $\vec k'$. In this case, for the off-mass-shell virtual states such that $v_F|\vec p|\gg \mu,\, k_{\rm B}T$, we can approximate $G(\vec p)\approx -\frac1{H_0(\vec p)}$. The stiffness of electron dispersion, $v_F\gg s$, along with the estimate $|\vec p|\sim q_T$, makes it an accurate approximation for all virtual states not too close to the Fermi surface. In this limit, as we now show, drastic simplifications occur because of the particle-hole symmetry $H_0(-\vec p)=-H_0(\vec p)$.

We focus on the case of short-range disorder, modeled by a delta function potential $V(\vec r-\vec r_j)=u\delta(\vec r-\vec r_j)(\hat 1\pm\sigma_z)/2$, where the plus (minus) sign correspond to impurity positions on the A (B) sites of the carbon lattice. In this case, a nonzero result for the transition matrix element $M_\pm$ is obtained at first order in $u$. We approximate $\hat T_{\vec p',\vec p}=\frac12 u (\hat 1\pm\sigma_z) e^{i(\vec p'-\vec p)\vec r_j}+O(u^2)$ and evaluate the first two terms in Eq.(\ref{matrixelement_M}). This gives the commutator of $H_0^{-1}(\vec q)$ and $\pm\sigma_z$, arising because the virtual electron states in the first and second term have momenta $\vec p\approx -\vec q$ and $\vec p\approx +\vec q$ (see above). We obtain
\be\label{eq:M1}
M_\pm = \frac{\pm i u g\sqrt{\omega_{\vec q}}}{\hbar v_F|\vec q|^2}\la \vec k'|(\sigma \times\vec q)_z |\vec k\ra e^{i(\vec p'-\vec p \pm \vec q)\vec r_j}
,
\ee
with the phase factor describing the dependence on the impurity position.
We evaluate the energy-loss power as
\be
\mathcal{J}= \sum_{\vec k, \vec{k'}, i} W_{\vec{k}', \vec{k}} (\epsilon_\vec{k} - \epsilon_\vec{k'}) f(\epsilon_\vec{k}) \big[ 1- f(\epsilon_\vec{k'})\big]  
\label{cool}
\ee
where $f(\epsilon) = 1/(e^{\beta(\epsilon -\mu)} + 1)$ are Fermi functions, $W_{\vec{k}', \vec{k}}$ is the transition probability, and $\epsilon_\vec{k} - \epsilon_\vec{k'}$ is the energy exchanged in each scattering event.  

In the degenerate limit, $k_BT\ll \mu$, the sum over $\vec k$ and $\vec{k'}$ is conveniently factored into separate integration over energies and angles $ \sum_{\vec{k},\vec{k'}}= \lb\nu(\mu) \rb^2 \int\!\!\int d\epsilon d\epsilon' \int\!\!\int \frac{d\theta d\theta'}{(2\pi)^2}$.
One of the energy integrals is eliminated by the delta functions $\delta(\epsilon_\vec{k'} - \epsilon_\vec{k}\pm\omega_{\vec q})$. The second energy integral is evaluated using the identity 
$
\int_{-\infty}^{\infty} d\epsilon f(\epsilon) (1- f(\epsilon + \omega_{q})) = \omega_{\vec q} \big[N^{\rm el}_{\omega_{\vec q}}+1 \big]
$, 
where $N^{\rm el}$ is the Bose distribution function evaluated at the {\it electron temperature}.
With the electron-phonon matrix element given by Eq.(\ref{eq:M1}), and using the angle-averaged quantity $\la|\la \vec k'|(\sigma\times\vec q)_z|\vec k\ra|^2\ra_{\rm ave}= |\vec q|^2/2$
 we obtain an expression
\be\nonumber
\mathcal{J}= \frac{\pi Ng^2 u^2}{\hbar^3 v_F^2} \lb\nu(\mu)\rb^2 n_0\sum_{\vec q}
\frac{\omega_{\vec q}^3}{|\vec q|^2}\lb N^{\rm el}_{\omega_{\vec q}} - N_{\omega_{\vec q}} \rb
,
\ee
where $n_0$ is impurity concentration. 
Integration yields Eq.(\ref{eq:Jsuper}), where we used 
an expression for the mean free path $k_F\ell=2\hbar^2 v_F^2/( u^2 n_0)$\cite{ando}.

We can make a comparison with the normal (momentum conserving) processes\cite{macdonald,wong}, where the cooling power is $\mathcal{J}_0 = B(T_{\rm el}-T_{\rm ph})$, where $B=\pi N \lambda \hbar \nu(\mu)  k_F^2 s^2k_{\rm B}$, and $\lambda=g^2\nu(\mu)$ is the dimensionless electron-phonon coupling. Linearizing Eq.(\ref{eq:Jsuper}) in $\Delta T=T_{\rm el}-T_{\rm ph}$, we find that this contribution dominates over $\mathcal{J}_0 $ at temperatures
\be
T>T_*=\sqrt{\frac{B}{3A}}=\lp \frac{\pi}{6\zeta(3)} k_F\ell\rp^{1/2} T_{\rm BG}
.
\ee
Taking $k_F\ell=20$ for a rough estimate, we see that the disorder assisted cooling channel dominates for $T\gtrsim 3T_{\rm BG}$. The crossover temperature  can be controlled by gate voltage, since $T_{\rm BG}\propto \sqrt{n}$. For typical carrier densities $n$ this gives a crossover temperature $T_*$ of a few tens of kelvin.

Interestingly, the cooling times describing relaxation to equilibrium, $\Delta T_{\rm el} (t) =   e^{- (t-t_0)/\tau}\Delta T_{\rm el,0}$, exhibit a nonmonotonic $T$ dependence for $T\sim T_*$. Accounting for both the disorder-assisted and momentum-conserving cooling, the relaxation dynamics can be described as 
\be\label{eq:dQ/dt}
d\mathcal{Q}/dt=-\mathcal{J} -\mathcal{J}_0
,
\ee
where $\mathcal{Q}$ is the electron energy density. Taking $\mathcal{Q}=C\Delta T$, with  $C=\alpha T$ the heat capacity of the degenerate electronic system ($\alpha=\frac{\pi^2}3N\nu(\mu)k_{\rm B}^2$), we find
\be
\label{eq:nonmono}
\frac{1}{\tau}=\frac{3A}{\alpha}T+\frac{B}{\alpha T}.
\ee
The cooling time increases with $T$ at $T<T_*$ and decreases at $T>T_*$, reaching maximal value at $T=T_*$. The non-monotonic temperature dependence provides a clear experimental signature of the competition between different cooling pathways.

To describe the cooling dynamics both near and away from equilibrium, we used non-linearized quantities, $\mathcal{Q}=\frac12\alpha T_{\rm el}^2$ and Eq.(\ref{eq:Jsuper}), with the deformation potential constant $D=20\,{\rm eV}$\cite{wong,macdonald}, the electron temperature initial value $T_{\rm el,0}=3\cdot 10^3\,{\rm K}\sim\omega_0$ and other parameter values cited in Fig.\ref{compare} caption. 
For the parameters used,
$T_* \approx 15 \, {\rm K}$. The resulting dynamics, shown in Fig.\ref{compare}(a), reproduces the main features seen in the data. We note that the long time behavior is insensitive to the choice of $T_{\rm el,0}$; only the dynamics at short times are affected. 

The non-exponential behavior seen in the data at short times can be understood by analyzing the regime $T_{\rm el} \gg T_{\rm ph}$. Approximating $\mathcal{J}\approx AT_{\rm el}^3$ and suppressing $\mathcal{J}_0$, we obtain a $1/(t-t_0)$ dynamics:
\be\label{eq:1/t}
T_{\rm el}(t)
=\frac{T_{\rm el,0}}{1+(A/\alpha)(t-t_0)T_{\rm el,0}}
.
\ee
The dynamics at intermediate times, where $ T_{\rm el} \gtrsim T_{\rm ph}$, can be found by directly solving Eq.(\ref{eq:dQ/dt}).
We obtain
\be
\label{eq:fulldynamics}
-\frac{2}{\tau} (t-t_0) = F(T_{\rm el}(t)/ T_{\rm ph}) - F(T_{\rm el,0}/T_{\rm ph}),
\ee  
where $F(x) = 2\sqrt{3} \arctan [(1+2x)/\sqrt{3}] - {\rm ln} [ (x^3-1)/(x-1)^3]$ (we suppressed the $\mathcal{J}_0$ term which becomes important only for $T\lesssim T_*$ and only at long times). 
This solution, with $\tau$ taken from Eq.(\ref{eq:nonmono}),  was used to generate Fig. \ref{compare} (a),
yielding results strikingly similar to the data. 

Interestingly, both at $T<T_*$ and $T> T_*$
the supercollision frequency remains lower than that for normal processes, 
$\la W_0\ra=\frac{2\pi }{\hbar} \lambda k_{\rm B}T$\cite{macdonald,wong}. 
We can define the average collision frequency as
\be
\la W\ra =\frac{\sum_{\vec k, \vec{k'}} W_{\vec{k}', \vec{k}} f(\epsilon_\vec{k}) \big[ 1- f(\epsilon_\vec{k'})\big]  }{\sum_{\vec k} f(\epsilon_\vec{k}) \big[ 1- f(\epsilon_\vec{k})\big]}
.
\ee
Evaluating the
integrals as above and setting $T_{\rm el}=T_{\rm ph}$, we find $\la W\ra = \frac{4\pi}{\hbar k_F\ell} \lambda k_{\rm B}T\ln\frac{T}{T_{\rm BG}} \ll \la W_0\ra$. 
The low value for $\la W\ra$ means that normal collisions produce the dominant contribution to resistivity even when their contribution to cooling is totally overwhelmed by supercollisions.

Finally, we analyze cooling in free standing graphene in the absence of disorder. In this case, an important contribution arises due to flexural phonons\cite{castro,ochoa}, which contribute to the deformation tensor via $u_{ij} = 1/2 (\partial_i u_j + \partial_j u_i + \partial_i h\partial_j h)$, with $\vec{u}$ and $h$ the in-plane and out-of-plane displacements. Flexural modes have quadratic dispersion $\tilde\omega_\vec{q} =  \kappa |\vec q|^2$ with
$\kappa\approx 4.6 \times 10^{-7} \, {\rm m}^2 {\rm s}^{-1}$\cite{castro,ochoa}. Electron-phonon coupling is described by the same deformation potential as above, Eq.(\ref{eq:H_elph}).

The processes involving pairs of nearly counterpropagating flexural phonons are analyzed as follows.
Using the momentum representation, $h_\vec{q} = \sqrt{\hbar/2\rho \tilde\omega_q }\big(b_\vec{q} + b_{-\vec q}^\dag\big)$, 
we consider the emission/absorption of two flexural phonons with momenta $\vec{q_1}$ and $\vec{q_2}$. For $T \gg T_{\rm BG}^{\rm flex} = \hbar \kappa k_F^2$ (for typical densities, $T_{\rm BG}^{\rm flex}$ is well below $1\,{\rm K}$),  we can set $\vec{q_1} \approx - \vec{q_2} = \vec{q}$, yielding the transition probability 
\be
W_{\vec k', \vec k} \!\!=   \!\! \frac{2\pi}{\hbar} \sum_\vec{q} |\la \vec{k'}|\vec k\ra|^2  M ^2 \Big[ N_{\omega_{\vec q}}^2  \delta_+ + (N_{\omega_{\vec q}}+1)^2 \delta_-\Big]
, 
\ee
$\delta_\pm = \delta( \epsilon_{\vec k}-\epsilon_{\vec k'}\pm 2\tilde\omega_\vec{q})$. Here the matrix element is $M = \frac{D\hbar}{4 \rho \kappa}$ \cite{castro,ochoa} and the coherence factor is $|\la \vec{k'}|\vec k\ra|^2  = \big[1 \pm {\rm cos} ( \theta_\vec{k} - \theta_\vec{k'} )\big]/2$, with the plus (minus) sign for intra-band (inter-band) processes. This gives the energy-loss power 
\[
\mathcal{J}_{\rm flex} = \sum_q
%\frac{\pi N D^2\hbar^3 }{16 \rho^2 \kappa^2} \big[\nu(\mu) \big]^2 
%LL C \int\!\! \frac{d^2q}{(2\pi)^2} 
(2\hbar\tilde\omega_q)^2  \lb (N_{\omega_{\vec q}} \!\!+ 1)^2 N^{\rm el}_{2\omega_{\vec q}} -N_{\omega_{\vec q}}^2 ( N^{\rm el}_{2\omega_{\vec q}} \!\!+1)\rb
% \Delta,
\]
%
%LL $C=\frac{\pi N D^2\hbar^2 }{16 \rho^2 \kappa^2} \big[\nu(\mu) \big]^2 $.
where $\sum_q...=\frac{\pi N D^2\hbar^2 }{16 \rho^2 \kappa^2} \big[\nu(\mu) \big]^2 \int\!\! \frac{d^2q}{(2\pi)^2}...$
We note that the above expression
%LL  in Eq.(\ref{eq:2F_cooling_power}) 
vanishes when $T_{\rm el} = T_{\rm ph}$.

We linearize $\mathcal{J}_{\rm flex}$ in $\Delta T = T_{\rm el} - T_{\rm ph}$ to obtain
\be\label{eq:2F_cooling_power}
\mathcal{J}_{\rm flex} = A_1 T^2 \Delta T, \qquad A_1 = 0.12\frac{ N D^2 \nu^2(\mu)k_{\rm B}^3}{\rho^2 \kappa^3},
\ee
which scales with $T$ the same way as Eq.(\ref{eq:Jsuper}). 
Flexural phonons dominate over the one-phonon contribution at
\be
T > T_*^{\rm flex}=\sqrt{\frac{B}{A_1}}=\Big( \frac{\pi  \rho \kappa^3}{0.24 \hbar s^2}   \Big)^{1/2} T_{\rm BG}  \approx 10 \,T_{\rm BG}
.
\ee
The value $T_*^{\rm flex}$ is considerably larger than $T_*$ for disorder-assisted cooling estimated above. A comparison with Eq(\ref{eq:Jsuper}) yields $\mathcal{J}_{\rm flex}/ \mathcal{J} \approx k_F \ell/ 200$ which is small for typical $k_F\ell$.Thus the contribution (\ref{eq:2F_cooling_power}) is relatively weak under realistic conditions.
For graphene on substrate this contribution is further diminished as flexural modes get pinned, gapped, and stiffened by the substrate.

Besides carrier dynamics, cooling can also be probed by transport measurements through bolometry \cite{fuhrer}, described by the thermal impedance, $R_{\rm th} = (d\mathcal{P}/d\Delta T)^{-1}$. Here $\mathcal{P}$ is the power pumped into the system (say, via Joule heating),
which is balanced by $\mathcal{J}$ in a steady state.
Hence, $R_{\rm th}$ temperature dependence can be used as a diagnostic for the processes dominant in cooling. For disorder-assisted cooling, extending Eq.(\ref{eq:Jsuper}) to $\mu\lesssim k_{\rm B}T$, we approximate $R_{\rm th} ^{-1}= AT^{2}\big( 1+ c (k_{\rm B}T/\mu)^2 \big)$ for monolayer graphene, with $c$ a constant of order unity. 
This temperature dependence is markedly different from $R_{\rm th}$ predicted for momentum conserving channels\cite{viljas}.
 
In summary, graphene stands out as a unique system where disorder-limited cooling is the {\it leading} contribution over a wide range of temperatures, including room temperature. 
%We have argued that disorder-assisted cooling represents the {\it limiting} process for cooling in a range of experimentally relevant temperatures, including room temperature. 
As a result, varying the amount of disorder can be used as a knob to tailor and control a variety of hot carrier effects in graphene. Tuning disorder can be achieved by well established techniques, including current-annealing and using different substrates (e.g. SiO2 or BN). The characteristic $T^3$ dependence, Eq.(\ref{eq:Jsuper}), and power-law cooling dynamics, Eq.(\ref{eq:1/t}), makes this new regime easy to identify in experiments\cite{mceuen,xuprivate}. %We anticipate that tuning scattering rates with disorder can be used as a knob to tailor and control a variety of hot carrier effects in graphene.}

We acknowledge useful discussions with P. Kim, F. Rana, M. Serbyn, A. Sergeev, A. Shytov, B.Z. Spivak, and support from the NSS program, Singapore (JS) and the Office of Naval Research
Grant N00014-09-1-0724 (LL).

% \end{document}
% \newpage

% \newpage

\end{document}